\documentclass[11pt]{article}
\usepackage{moriond,epsfig}

\def\Journal#1#2#3#4{{#1} {\bf #2}, #3 (#4)}

\def\PLB{{\em Phys. Lett.}  B}
\def\PRL{\em Phys. Rev. Lett.}
\def\PRD{{\em Phys. Rev.} D}

\def\be{\begin{equation}}
\def\ee{\end{equation}}
\def\bea{\begin{eqnarray}}
\def\eea{\end{eqnarray}}

\begin{document}
\vspace*{4cm}
\title{A NEW LIGHT BOSON FROM CHERENKOV TELESCOPES OBSERVATIONS?}

\author{MARCO RONCADELLI }

\address{INFN, Sezione di Pavia, via A. Bassi 6, I -- 27100 Pavia, Italy}

\author{ALESSANDRO DE ANGELIS}

\address{Dipartimento di Fisica, Universit\`a di Udine, Via delle Scienze 208, I -- 33100 Udine,\\ 
and INAF and INFN, Sezioni di Trieste, Italy}

\author{ORIANA MANSUTTI}

\address{Dipartimento di Fisica, Universit\`a di Udine, Via delle Scienze 208, I -- 33100 Udine,\\ 
and INFN, Sezione di Trieste, Italy}

\maketitle\abstracts{Early indications by H.E.S.S. and the subsequent detection of blazar 3C279 by MAGIC show that the Universe is more transparent to 
very-high-energy gamma rays than previously thought. We demonstrate that this circumstance can be reconciled with standard blazar emission models provided that photon oscillations into a very light Axion-Like Particle occur in extragalactic magnetic fields. A quantitative estimate of this effect indeed explains the observed spectrum of 3C279. Our prediction can be tested by the satellite-borne {\it Fermi}/LAT detector as well as by the ground-based Imaging Atmospheric Cherenkov Telescopes H.E.S.S., MAGIC, CANGAROO III, VERITAS and by the Extensive Air Shower arrays ARGO-YBJ and MILAGRO.}

\section{Introduction}
A characteristic feature of the very-high-energy (VHE) band is that the horizon of the observable Universe rapidly shrinks above $100 \, {\rm GeV}$ as the energy further increases. This is due to the fact that photons from distant sources scatter off background radiation permeating the Universe, thereby disappearing into electron-positron pairs~\cite{stecker1971}. The corresponding cross section $\sigma (\gamma \gamma \to e^+ e^-)$ peaks where the VHE photon energy $E$ and the background photon energy $\epsilon$ are related by $\epsilon \simeq (500 \, {\rm GeV}/E) \, {\rm eV}$. We recall that Imaging Atmospheric Cherenkov Telescopes (IACTs) probe the energy interval $100 \, {\rm GeV} - 100 \,{\rm TeV}$. Consequently, observations performed by the IACTs are affected by an opacity dominated by the interaction of the beam photon with ultraviolet/optical/infrared diffuse background photons (frequency band $1.2 \cdot 10^{3} \, {\rm GHz} - 1.2 \cdot 10^{6} \, {\rm GHz}$, corresponding to the wavelength range $0.25 \, \mu {\rm m} - 250 \, \mu {\rm m}$), usually called Extragalactic Background Light (EBL) and produced by galaxies during the whole history of the Universe. Neglecting evolutionary effects for simplicity, photon propagation is then controlled by the photon mean free path ${\lambda}_{\gamma}(E)$ for $\gamma \gamma \to e^+ e^-$, and so the observed photon spectrum $\Phi_{\rm obs}(E,D)$ is related to the emitted one $\Phi_{\rm em}(E)$ by 
\begin{equation}
\label{a1}
\Phi_{\rm obs}(E,D) = e^{- D/{\lambda}_{\gamma}(E)} \ \Phi_{\rm em}(E)~.
\end{equation}

Within the considered energy range, ${\lambda}_{\gamma}(E)$ decreases like a power law from the Hubble radius $4.3 \, {\rm Gpc}$ around $100 \, 
{\rm GeV}$ to nearly $1 \, {\rm Mpc}$ around $100 \, {\rm TeV}$~\cite{CoppiAharonian}. Thus, Eq.~(\ref{a1}) implies that the observed flux is {\it exponentially} suppressed both at high energies and at large distances, so that sufficiently far-away sources become hardly visible in the VHE range and their observed spectrum should anyway be {\it much steeper} than the emitted one.

Yet, the behaviour predicted by Eq.~(\ref{a1}) has {\it not} been detected by observations. A first indication in this direction was reported by the H.E.S.S. collaboration in connection with the discovery of the two blazars H2356-309 ($z = 0.165$) and 1ES1101-232 ($z = 0.186$) at $E \sim 1 \, {\rm TeV}$~\cite{aharonian:nature06}. Stronger evidence comes from the observation of blazar 3C279 ($z = 0.536$) at $E \sim 0.5 \, {\rm TeV}$ by the MAGIC collaboration~\cite{3c}. In particular, the signal from 3C279 collected by MAGIC in the region $E<220$ GeV has more or less the same statistical significance as the one in the range 220 GeV $< E <$ 600 GeV ($6.1 \sigma$ in the former case, $5.1 \sigma$ in the latter)~\footnote{See ref.~\cite{dis} for a different view.}. 

A possible way out of this difficulty involves the modification of the standard Synchro-Self-Compton (SSC) emission mechanism. One option invokes strong relativistic shocks~\cite{Stecker2007}. Another is based on photon absorption inside the blazar~\cite{Aharonian2008}. While successful at substantially hardening the emission spectrum, these attempts fail to explain why {\it only} for the most distant blazars does such a drastic departure from the SSC emission spectrum show up.

Our proposal -- usually referred to as the DARMA scenario -- is quite different~\cite{drm}. Implicit in all previous considerations is the hypothesis that photons propagate in the standard way throughout cosmological distances. We suppose instead that photons can oscillate into a new very light spin-zero particle -- named Axion-Like Parlicle (ALP) -- and vice-versa in the presence of cosmic magnetic fields, whose existence has definitely  been proved by AUGER observations~\cite{auger}. Once ALPs are produced close enough to the source, they travel {\it unimpeded} throughout the Universe and can convert back to photons before reaching the Earth. Since ALPs do not undergo EBL absorption, the  {\it effective} photon mean free path ${\lambda}_{\gamma , {\rm eff}} (E)$ gets {\it increased} so that the observed photons travel a distance in excess of ${\lambda}_{\gamma}(E)$. Correspondingly, Eq. (\ref{a1}) becomes
\begin{equation}
\label{a1bis}
\Phi_{\rm obs}(E,D) = e^{- D/{\lambda}_{\gamma , {\rm eff}}(E)} \ \Phi_{\rm em}(E)~,
\end{equation}
which shows that even a {\it small} increase of ${\lambda}_{\gamma , {\rm eff}}(E)$ gives rise to a {\it large} enhancement of the observed flux. It turns out that the DARMA mechanism makes ${\lambda}_{\gamma , {\rm eff}}(E)$ shallower than ${\lambda}_{\gamma}(E)$ although it remains a decreasing function of 
$E$. So, the resulting observed spectrum is {\it much harder} than the one predicted by Eq. (\ref{a1}), thereby ensuring agreement with observations even for a {\it standard} SSC  emission spectrum. As a bonus, we get a natural explanation for the fact that only the most distant blazars would demand $\Phi_{\rm em}(E)$ to substantially depart from the emission spectrum predicted by the SSC mechanism.

We proceed to review the main features of our proposal as well as its application to blazar 3C279.

\section{DARMA scenario}

Both phenomenological and conceptual arguments entail that the Standard Model (SM) of particle physics should be viewed as the low-energy manifestation of some more fundamental and richer theory of all elementary-particle interactions including gravity. Therefore, the SM lagrangian is expected to be modified by small terms describing interactions among known and new particles. Many extensions of the SM which have attracted considerable interest in the last few years indeed predict the existence of ALPs. They are spin-zero light bosons defined by the low-energy effective lagrangian
\begin{equation}
\label{a1a}
{\cal L}_{\rm ALP} \ = \ 
\frac{1}{2} \, \partial^{\mu} \, a \, \partial_{\mu} \, a - \frac{1}{2} 
\, m^2 \, a^2 - \frac{1}{4 M} \, F^{\mu \nu} \, \tilde F_{\mu \nu} \, a~,
\end{equation}
where $F^{\mu \nu}$ is the electromagnetic field strength, $\tilde F_{\mu \nu}$ is its dual, $a$ denotes the ALP field and $m$ stands for the ALP mass. According to the above view, it is assumed $M \gg G_F^{- 1/2} \simeq 250 \, {\rm GeV}$. On the other hand, it is supposed that $m \ll G_F^{- 1/2} 
\simeq 250 \, {\rm GeV}$. The standard Axion~\cite{axion} is the archetype of ALPs and is characterized by a specific relation between $M$ and $m$, while in the case of {\it generic} ALPs $M$ and $m$ are to be regarded as {\it independent}. So, the peculiar feature of ALPs is the trilinear $\gamma$-$\gamma$-$a$ vertex described by the last term in ${\cal L}_{\rm ALP}$, whereby one ALP couples to two photons. 

Owing to such a vertex, ALPs can be emitted by astronomical objects of various kinds, and this fact yields strong bounds: $M > 0.86 \cdot 10^{10} \, {\rm GeV}$ for $m < 0.02 \, {\rm eV}$~\cite{Zioutas2005} and  $M > 10^{11} \, {\rm GeV}$ for $m < 10^{- 10} \, {\rm eV}$~\cite{Raffelt1990}. Moreover, the same $\gamma$-$\gamma$-$a$ vertex produces an off-diagonal element in the mass matrix for the photon-ALP system in the presence of an external magnetic field ${\bf B}$. Therefore, the interaction eigenstates differ from the propagation eigenstates and photon-ALP oscillations show up~\cite{Sikivie1984}.

We imagine that a sizeable fraction of photons emitted by a blazar soon convert into ALPs. They propagate unaffected by the EBL and we suppose that before reaching the Earth a substantial fraction of ALPs is back converted into photons. We further assume that this photon-ALP oscillation process is triggered by cosmic magnetic fields (CMFs), whose existence has been demonstrated very recently by AUGER observations~\cite{auger}. Owing to the notorious lack of information about their morphology, one usually supposes that CMFs have a domain-like structure~\cite{Kronberg}. That is, ${\bf B}$ ought to be constant over 
a domain of size $L_{\rm dom}$ equal to its coherence length, with ${\bf B}$ randomly changing its direction from one domain to another but keeping approximately the same strength. As explained elsewhere~\cite{dpr}, it looks plausible to assume the coherence length in the range $1 - 10 \, {\rm Mpc}$. Correspondingly, the inferred strength lies in the range $0.3 - 1.0 \, {\rm nG}$. 

\section{Predicted energy spectrum}

Our ultimate goal consists in the evaluation of the probability $P_{\gamma \to \gamma}(E,D)$ that a photon remains a photon after propagation from the source to us when allowance is made for photon-ALP oscillations as well as for photon absorption from the EBL. As a consequence, Eq. (\ref{a1bis}) gets replaced by
\begin{equation}
\label{a0as}
\Phi_{\rm obs}(E,D) = P_{\gamma \to \gamma}(E,D) \, \Phi_{\rm em}(E)~. 
\end{equation}
Our procedure is as follows. We first solve exactly the beam propagation equation arising from ${\cal L}_{\rm ALP}$ over a single domain, assuming that the EBL is described by the ``best-fit model'' of Kneiske {\it et al.}~\cite{kneiske}. Starting with an unpolarized photon beam, we next propagate it by iterating the single-domain solution as many times as the number of domains crossed by the beam, taking each time a {\it random} value for the angle between ${\bf B}$ and a fixed overall fiducial direction. We repeat such a procedure $10^.000$ times and finally we average over all these realizations of the propagation process. 

We find that about 13\% of the photons arrive to the Earth for $E = 500 \, {\rm GeV}$, representing an enhancement by a factor of about 20 with respect to the expected flux without DARMA mechanism (the comparison is made with the above ``best-fit model''). The same calculation gives a fraction of 76\% for $E = 100 \, {\rm GeV}$ (to be compared to 67\% without DARMA mechanism) and a fraction of 3.4\% for $E =  1 \, {\rm TeV}$ (to be compared to 0.0045\% without DARMA mechanism). The resulting spectrum is exhibited in Fig.~1. The solid line represents the prediction of the DARMA scenario for  \mbox{$B \simeq 1 \, {\rm nG}$} and \mbox{$L_{\rm dom} \simeq 1 \, {\rm Mpc}$} and the gray band is the envelope of the results obtained by independently varying ${\bf B}$ and $L_{\rm dom}$ within a factor of 10 about such values. These conclusions hold for $m < 10^{-10} \, {\rm eV}$ and we have taken for definiteness $M \simeq 4 \cdot 10^{11} \, {\rm GeV}$ but we have cheked that practically nothing changes for $10^{11} \, {\rm GeV} < M < 10^{13} \, {\rm GeV}$.

Our predictions can be tested by the satellite-borne {\it Fermi}/LAT detector as well as by the ground-based IACTs H.E.S.S., MAGIC, CANGAROO III, VERITAS and by the Extensive Air Shower arrays ARGO-YBJ and MILAGRO.

\begin{figure}[hb]
\centerline{\includegraphics[width=0.76\textwidth]{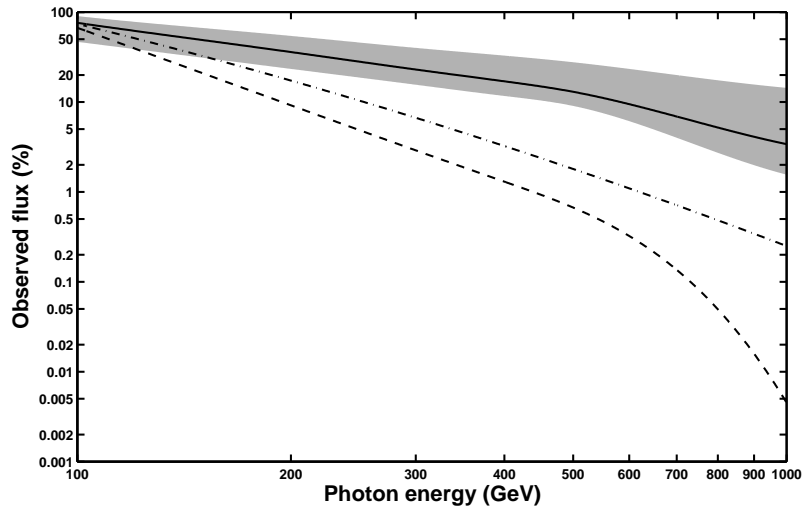}}
\caption{The two lowest lines give the fraction of photons surviving from 3C279 without the DARMA mechanism within the ``best-fit model'' of EBL (dashed line) and for the minimum EBL density compatible with cosmology (dashed-dotted line), which are discussed by Kneiske {\it et al.} (see reference in the text). The solid line represents the prediction of the DARMA mechanism.} \label{Fig:MV}
\end{figure}

\end{document}